\documentclass[aps,prl,twocolumn,showpacs,preprintnumbers,amsmath,amssymb,superscriptaddress]{revtex4-1}
\usepackage{color}
\usepackage{graphicx}
\usepackage{dcolumn}
\usepackage{bm}

\usepackage[hypertex]{hyperref}
\newcommand{\nc}{\newcommand}
\nc{\be}{\begin{eqnarray}}
\nc{\ee}{\end{eqnarray}}
\nc{\bea}{\begin{eqnarray}}
\nc{\eea}{\end{eqnarray}}
\nc{\bean}{\begin{eqnarray*}}
\nc{\eean}{\end{eqnarray*}}
\nc{\mb}{\mbox}
\nc{\rnc}{\renewcommand}
\nc{\vk}{\mb{\boldmath$k$}}
\nc{\vx}{\mb{\bf x}}
\nc{\br}{\mb{\bf r}}
\nc{\bv}{\mb{\bf v}}
\nc{\bp}{\mb{\bf p}}
\nc{\ve}{\mb{\bf e}}
\nc{\vz}{\hat {\mb{\bf z}}}
\nc{\vp}{\mb{\boldmath$p$}}
\nc{\vb}{\mb{\boldmath$b$}}
\nc{\rr}{\mb{\boldmath$r$}}
\nc{\vR}{\mb{\boldmath$R$}}
\nc{\vj}{\mb{\boldmath$j$}}
\nc{\vg}{\mb{\boldmath$g$}}
\nc{\vm}{\mb{\boldmath$m$}}
\nc{\vd}{\mb{\boldmath$d$}}
\nc{\hd}{\mb{\boldmath$\hat{d}$}}
\nc{\vD}{\mb{\boldmath$D$}}
\nc{\vF}{\mb{\boldmath$F$}}
\nc{\vG}{\mb{\boldmath$G$}}
\nc{\vI}{\mb{\boldmath$I$}}
\nc{\vW}{\mb{\boldmath$W$}}
\nc{\x}{\mb{\boldmath$x$}}
\nc{\A}{\mb{\boldmath$A$}}
\nc{\va}{\mb{\boldmath$a$}}
\nc{\vv}{\mb{\boldmath$v$}}
\nc{\vq}{\mb{\boldmath$q$}}
\nc{\vn}{\mb{\boldmath$n$}}
\nc{\vJ}{\mb{\boldmath$J$}}
\nc{\vS}{\mb{\boldmath$S$}}
\nc{\vs}{\mb{\boldmath$\sigma$}}
\nc{\vE}{\mb{\boldmath$E$}}
\nc{\vB}{\mb{\boldmath$B$}}
\nc{\vM}{\mb{\boldmath$M$}}
\nc{\vL}{\mb{\boldmath$L$}}
\nc{\vpsi}{\mb{\boldmath$\psi$}}
\nc{\vphi}{\mb{\boldmath$\varphi$}}
\nc{\Vphi}{\mb{\boldmath$\phi$}}
\nc{\Vomega}{\mb{\boldmath$\Omega$}}
\nc{\ipsi}{\it{\Psi}}
\nc{\vepsilon}{\mb{\boldmath$\epsilon$}}
\nc{\valpha}{\mb{\boldmath$\alpha$}}
\nc{\vgamma}{\mb{\boldmath$\gamma$}}
\nc{\vomega}{\mb{\boldmath$\omega$}}
\nc{\vmu}{\mb{\boldmath$\mu$}}
\nc{\vt}{\mb{\boldmath$\tau$}}
\nc{\vT}{\mb{\boldmath$T$}}
\nc{\vpi}{\mb{\boldmath$\pi$}}
\nc{\nab}{\nabla}
\nc{\ov}{\overline}
\nc{\cdott}{\!\cdot\!}
\nc{\cdottt}{\!\!\cdot\!}
\nc{\LL}{\Big{\langle}}
\nc{\RR}{\Big{\rangle}}
\nc{\LR}{\Bigm{|}}
\nc{\vP}{\mb{\boldmath$P$}}
\nc{\nnn}{\nonumber\\}

\begin{document}

\title{
Cross-Correlated Responses of Topological Superconductors and Superfluids
}

\author{Kentaro Nomura}
\affiliation{
Correlated Electron Research Group (CERG), RIKEN-ASI, Wako 351-0198, Japan}
\author{Shinsei Ryu}
\affiliation{
Department of Physics, University of Illinois, 1110 West Green St, Urbana IL 61801, USA}
\affiliation{
Condensed Matter Theory Laboratory, RIKEN, Wako, Saitama 351-0198, Japan}     
\author{Akira Furusaki}
\affiliation{
Condensed Matter Theory Laboratory, RIKEN, Wako, Saitama 351-0198, Japan}
\author{Naoto Nagaosa}
\affiliation{
Correlated Electron Research Group (CERG), RIKEN-ASI, Wako 351-0198, Japan}
\affiliation{
Cross-Correlated Material Research Group (CMRG), RIKEN-ASI, Wako 351-0198, Japan}
\affiliation{
Department of Applied Physics, The University of Tokyo, Hongo, Bunkyo-ku, Tokyo 113-8656, Japan
}

\date{\today}

\begin{abstract}
We study nontrivial responses of topological superconductors and
superfluids to the temperature gradient and rotation of the system.
In two-dimensional gapped systems, the Str\v{e}da formula for the electric
Hall conductivity is generalized to the thermal Hall conductivity.
Applying this formula to the Majorana surface states of
three-dimensional topological superconductors predicts
cross-correlated responses between the orbital angular momentum and
thermal polarization (entropy polarization).  These results can be
naturally related to the gravitoelectromagnetism description of
three-dimensional topological superconductors and superfluids,
analogous to the topological magnetoelectric effect in ${\mathbb{Z}}_2$
topological insulators.
\end{abstract}

\pacs{
73.43.-f, 74.25.fc, 74.90.+n, 74.25.F- 
}
\maketitle


\paragraph{Introduction}
The quantum Hall effect (QHE)
\cite{review_QHE} is a prominent example
of quantum phenomena characteristic
of insulators with topologically nontrivial electronic wave functions,
a class of materials called topological insulators (TIs)
\cite{review_TI}. In the QHE the Hall conductivity is quantized in
units of $e^2/h$ at integer values equal to
the topological number of bulk wave functions
\cite{TKNN}.  
The two-dimensional 
(2d) topological superconductors
(TSCs) and superfluids (TSFs)
with chiral ($p$-wave) Cooper pairing are
superconductor analogues of the QHE and considered to be realized,
e.g.,
in a thin film of $^3$He A phase
\cite{Leggetbook,Volovik}, Sr$_2$RuO$_4$
\cite{Maeno2003}, and the $\nu=5/2$
fractional QHE
\cite{Moore-Read}.
The topological nature of
such TSCs and TSFs
will manifest itself in thermal transport properties, such as
quantization of the thermal Hall conductivity
\cite{Read2000}.

Recent 
studies have 
shown that topological 
states exist in time-reversal invariant 
and 
three-dimensional
(3d)
cases
as well \cite{review_TI}, 
and
the systematic
classification of them is 
established
in terms of symmetries and dimensionality 
\cite{Schnyder2008,Kitaev2009}.
A key experimental signature of 
3d-TIs is
the topological magnetoelectric (ME) effect.  
Namely, the electromagnetic response of 3d-TIs 
is described
by the axion electrodynamics
\cite{review_TI,Qi2008,Essin2009}, 
\bea
 S_{\theta}^{\rm EM}=
\int dtd^3\x \,
\frac{e^2}{4\pi^2\hbar c}
\theta \vE\cdot\vB
\label{theta_EM}
\eea
with $\theta=\pi$, the possible nonzero value in time-reversal invariant
systems (mod $2\pi$).
The effective action (\ref{theta_EM}) leads to
the surface QHE
that induces the topological ME effect as
$\vM=(e^2/2hc)\vE$ and
$\vP=(e^2/2hc)\vB$
where 
$\vM$ and $\vP$ are
the magnetization and electric polarization, respectively.

\begin{figure}[b]
\begin{center}
\includegraphics[width=0.40\textwidth]{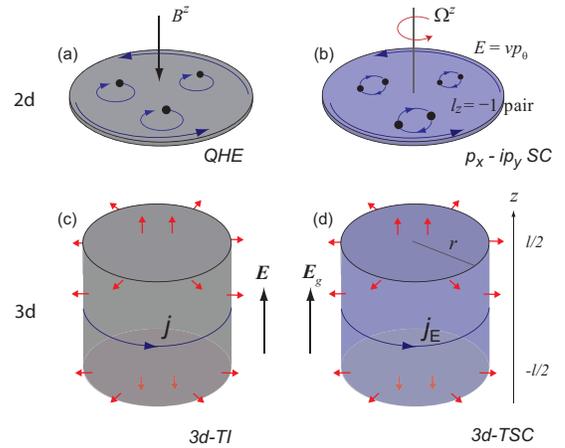}
\caption{(color online)
Electronic responses in 
(a)
two-dimensional (2d) and 
(c)
three-dimensional (3d)
topological insulators (TIs) (left)
and 
thermal and mechanical (rotating) responses in 
(b) 
2d and 
(d)
3d 
topological superconductors (TSCs) (right). 
In (b), the arrow along the boundary indicates
the chiral Majorana edge channel with dispersion $E=v p_{\theta}$.
In the 3d-TSC (d),
temperature gradient induces surface thermal Hall current $\boldsymbol{j}_E$.
A uniform mass gap is induced in the surface fermion spectrum by doping
magnetic impurities near the surface of the 3d-TI and TSC
such that spins are all pointing out or in (red arrows). 
\label{F1}
}
\end{center}
\end{figure}

\begin{table}[t]
\begin{ruledtabular}
\begin{tabular}{ccc}
& TI   &   TSC  \\
 \hline
\\
2d & 
$\displaystyle
\sigma_{H}^{}=ec\frac{\partial M^z}{\partial \mu}=ec\frac{\partial N}{\partial B^z}
$\ \ 
& 
$\displaystyle\kappa_{H}^{}=\frac{v^2}{2}\frac{\partial L^z}{\partial T}=\frac{v^2}{2}\frac{\partial S}{\partial \Omega^z}$ 
\\
\\
3d & $\displaystyle\chi_{\theta}^{ab}=\frac{\partial M^a}{\partial E^b} =\frac{\partial P^a}{\partial B^b}$ &
$\displaystyle
\chi^{ab}_{\theta,g}=\frac{\partial L^a}{\partial E^b_g}=\frac{\partial P^a_E}{\partial \Omega^b}$ 
\ \  \\
\\
        \end{tabular}
\end{ruledtabular}
\caption{
\label{table}
Comparison between
cross-correlation
in topological insulators 
(TIs)
and 
topological superconductors (TSCs)
in two (2d) and three spatial dimensions (3d). 
In TSCs
the orbital angular momentum $\vL$ and entropy $S$ 
(thermal polarization $\vP_E$ in 3d) 
are generated by temperature 
gradient ($\vE_g=-T^{-1}\nab T$) and by rotating the system with
angular velocity
$\Omega^{a}$.
In analogy with
the magnetoelectric polarizability
$\chi_{\theta}^{ab}$ in 3d TI
gravitomagnetoelectric polarizability
$\chi_{\theta,g}^{ab}$ can be introduced in 
the 3d TSC
(Right-bottom). 
Note that the relations 
for TSCs hold also 
for the thermal response of TIs.
These responses are characterized by 
the topological integers and quantized.
See Eq.\ (\ref{kappa2d}) for 2d-TSCs and Eq.\ (\ref{kappa-surface})
for the surface of 3d-TSCs}.
\end{table}

An example of 3d-TSFs \cite{TSF} is the B phase
of superfluid $^3$He \cite{Schnyder2008}. 
In addition, the newly found superconducting phase in Cu-doped Bi$_2$Se$_3$
\cite{Hor2010} 
has been proposed to be a 3d-TSC
\cite{Fu2010}.
The topological nature of 3d-TSCs
manifests itself in their coupling with 
the gravitational field,
which is described by the term similar to Eq.\ (\ref{theta_EM}),
the gravitational instanton term
\cite{Ryu2011,Wang2011}.

In this work we shall develop 
the response theory of TSCs,
which reveals the cross-correlation between
thermal and mechanical (rotational) responses,
and is the direct equivalent of
the topological ME effect in TIs.
We first generalize the Str\v{e}da formula 
\cite{Streda1982}
to the thermal Hall conductivity in 2d systems.
By applying it to the surface states of a 3d-TSC,
the cross-correlated responses of 3d-TSCs are identified with
those between
the orbital angular momentum and the thermal (entropy) polarization,
generated by temperature gradient $\nab T$ and
by rotating the system with 
angular velocity
$\Vomega$, respectively,
as shown in Fig.~\ref{F1}.
Our 
main
findings are summarized in 
Table \ref{table}. 
Their derivations will be given below.

\paragraph{Thermal Hall conductivity}

Relation between the Hall conductivity $\sigma_{H}^{}$ and
the magnetization $\vM$ in the quantum Hall regime is
known as the Str\v{e}da formula
\cite{Streda1982}:
\bea
 \sigma_{H}^{}=ec\frac{\partial M^z}{\partial\mu}
,
\label{streda-jc}
\eea
where $\mu$ is the chemical potential, 
$e$ ($<0$) is the electric charge,
and
$c$ is the speed of light. 
The magnetization is evaluated by
$
 \vM=(e/2c){\rm Tr}
[\theta(\mu-{\cal H})\x\times\vv],
$
where $\vv=(i/\hbar)[{\cal H},\x]$,
${\cal H}$ is the single-particle Hamiltonian,
and $\theta(x)$ is the step function.
The relation (\ref{streda-jc})
can be understood by identifying
the Hall current 
$\vj=\sigma_{H}^{}\vE\times\vz$
with 
$
\vj=c\nab\times\vM=-c\,\partial \vM/\partial\mu\times\nab\mu.
$

For the thermal conductivity
$\kappa_{H}^{}=j^x_T/\partial_yT$, 
we will show below that a similar relation holds:
\bea
 \kappa_{H}^{}=c\frac{\partial M^z_T}{\partial T}.
\label{streda-je}
\eea
The thermal current is $\vj_T=\vj_E-(\mu/e)\vj$, where 
$\vj$ is the electric charge current.
The energy-current $\vj_E$ is defined 
as $j_E^a=cT^{a0}$,
where $T^{a0}$ is a spatio-temporal component of
the energy-momentum tensor,
$
T^{\mu\nu}=
(\delta{\cal L}/\delta\partial_{\mu}\Psi) \partial^{\nu}\Psi
           -g^{\mu\nu}{\cal L},
$
satisfying the continuity equation, $\partial_{\mu}T^{\mu\nu}=0$;
${\cal L}[\Psi,\partial_{\mu}\Psi]$ is the Lagrangian density of the system.
The moment of the thermal current in Eq.\ (\ref{streda-je})
is defined as $\vM_T=\vM_E-(\mu/e)\vM$, where
$
M^{\mu\nu}_E=
\frac{1}{2}\left\langle x^{\mu}T^{\nu0}-x^{\nu}T^{\mu0}\right\rangle,
$
and $M^z_E=M^{12}_E$. 
In contrast to the charge Hall conductivity (\ref{streda-jc}),
the average has to be taken at finite temperature:
$\langle \cdots \rangle\equiv 
\sum_n f(\varepsilon_n)\langle n|\cdots|n\rangle$
where $\varepsilon_n$ and $|n\rangle$ are eigenvalue and eigenstate
of the Hamiltonian ${\cal H}$, 
and $f(\varepsilon_n)$ is the Fermi distribution function.
Below, we will consider the part of the thermal current 
carried by $\boldsymbol{j}_E$,  
specializing to the case of $\mu =0$.
For insulators, this means we 
count
the total number $N$ of 
particles
such that when the chemical potential is within a gap, $N=0$.
In the Bogoliubov-de Genne theory of superconductors,
due to particle-hole symmetry, this is always true, and hence 
$\boldsymbol{j}_T = \boldsymbol{j}_E$
and $\bm{M}_T=\bm{M}_E$.

Let us now
introduce a 
{\it gravitomagnetic}
field $\vB_g$ 
\cite{GEM}
which is conjugate to $\vM_E$ so that the
variation of free energy is 
$dF=-SdT-\vM_E\cdot d\vB_g$.  
Equation (\ref{streda-je}) is written as
$
 \kappa_{H}^{}=c(\partial M_E^z/\partial T)_{B^z_g}
 =c(\partial S/\partial B_g^z)_{T},
$
or
\bea
\kappa_{H}^{}
=\frac{c}{T}\left(\frac{\partial M^z_E}{\partial\phi}\right)_{B^z_g}
=\frac{c}{T}\left(\frac{\partial Q}{\partial B_g^z}\right)_\phi
\label{streda-je2}
\eea
by introducing $dQ=TdS$ and $d\phi=dT/T$,
where $\phi$ is a fictitious gravitational potential that couples
to
thermal energy density $Q$
\cite{Luttinger1964}.
Equation (\ref{streda-je2}) is 
 the thermal analogue of
the Str\v{e}da formula for the charge Hall conductivity,
$
 \sigma_{H}^{}=ec\,\partial M^z/\partial\mu=ec\,\partial N/\partial B^z
$, 
in that $Q$ is the zeroth component of the energy current 
as $eN$ is in the charge current.

To see the physical meaning 
of $\vB_g$ and $\vM_E$, 
note that the definition of $M_E^{\mu\nu}$ 
is similar to the orbital angular momentum:
$
 L^{\mu\nu}=
(1/c)   \left\langle x^{\mu}T^{0\nu}-x^{\nu}T^{0\mu}\right\rangle.
$
Indeed, 
when there is a relativistic invariance,
the energy-momentum tensor can be symmetrized so that $T^{\mu\nu}=T^{\nu\mu}$, 
and thus $M^{\mu\nu}_E= (c/2)L^{\mu\nu}$.
While this is not the case in condensed-matter 
systems in general,
in so-called pseudo-relativistic systems
where the Lorentz invariance is realized at low energies, 
electrons or quasiparticles obey the Dirac or Majorana equation.
In these systems the Fermi velocity $v$ 
plays a role of the speed of light $c$, 
and
the $M^{\mu\nu}_E$ tensor is related to the orbital angular momentum
as $\vM_E=(v/2)\vL$,
and 
$\vB_g$ can be understood as the angular velocity
vector of rotating systems,
$\vB_g=(2/v)\Vomega$.
For a 
system rotating with the frequency
$\boldsymbol{\Omega}=\Omega^z \vz$,
this can also be understood by making a coordinate transformation
from the rest frame to the rotating frame
in which the metric in the 
polar coordinates 
$(t,r,\varphi)$
takes the form 
$
 ds^2\simeq v^2dt^2-2 \Omega^z r^2d\varphi dt-r^2d\varphi^2-dr^2. 
$
One can then read off, from the definition of the 
gravitoelectromagnetic field, 
the non-zero gravito gauge potential 
$A_g^{\varphi} = \Omega^z r /v$ \cite{GEM}.
In Cartesian coordinates, 
$ \A_g=
(1/v)\Omega^z \vz\times\x,
$
and 
$\vB_g=\nab\times\A_g=(2/v)
\Omega^z \vz
$. 
Consequently,
$\kappa_{H}$ 
can be written as
\bea
\kappa_{H}^{}
=\frac{v^2}{2}\Bigg(\frac{\partial L^z}{\partial T} \Bigg)_{\Omega^z}
=\frac{v^2}{2}\Bigg(\frac{\partial S}{\partial \Omega^z}\Bigg)_{T}.
\label{streda-je3}
\eea
This is one of the main results of 
this work.

Although it is necessary to have (pseudo) Lorentz invariance to identify,
at the operator level, 
$\boldsymbol{M}_E$ with the angular momentum 
$\boldsymbol{L}$, 
the relation (\ref{streda-je3}) can in fact be derived for 
a wider range of systems,
e.g., by
assuming that the edge state 
in the disk geometry 
is described by chiral conformal field theory
\cite{note1}.
This indicates the validity of Eq.\ (\ref{streda-je3}) in
many 2d topological phases,
a representative example of which is
the 2d chiral $p$-wave SC with $l_z=-1$ pairing 
as shown in Fig.\ \ref{F1}(b). 
Near the edge, 
there exist chiral Majorana modes \cite{Read2000}
described by 
$H_{\rm edge}=(1/2) \int_0^L dx\, \psi(-i\hbar v \partial_x )\psi$,
where $\psi$ is a single-component real fermionic field and
$L$ is the circumference of the edge. 
Since
$T^{00}=T^{10}=T^{01}=(-i \hbar v/2)\psi \partial_x \psi$, 
$\langle j_E\rangle = (v/L)\langle H_{\rm edge}\rangle =
(v^2/2)\langle L^z_{\rm edge}\rangle= \pi^2k_B^2T^2/12h$,  
leading to 
\bea
\kappa_H^{}=\frac{\partial \langle j_E\rangle}{\partial T}=\frac{\pi^2k_B^2T}{6h}
\label{kappa2d}
,
\eea
which is a half of the quantized value for conventional chiral fermions.
Here,
$L^z_{\rm edge}$ is the edge modes contribution
to the total orbital angular momentum per unit area.

\paragraph{Majorana fermion on the surface of 3d-TSC/TSF}

Let us now derive 
Eqs.\ (\ref{streda-je2}-\ref{streda-je3})
from a microscopic theory
via Kubo formula.
We work with an example, 
the 2d massive Majorana fermion system
realized on the surface of 3d-TSCs
\cite{review_TI,Schnyder2008,Murakawa2011},
anticipating 
to apply 
(\ref{streda-je2}-\ref{streda-je3})
to derive the cross-correlated responses 
of the 3d-TSCs. 
It is described by 
$H=(1/2) \int d^2\x\, \psi^T\mathcal{H} \psi$, 
where  
$\psi^T=(\psi_{\uparrow},\psi_{\downarrow})$ is the real spinor field
satisfying
$\{\psi_{\alpha}(\x),\psi_{\beta}(\x')\}=\delta_{\alpha\beta}\delta(\x-\x')$,
and 
\bea
\mathcal{H}=
-i\hbar v (\sigma_z \partial_x +\sigma_x \partial_y) +m\sigma_y. 
\label{majorana-hamiltonian}
\eea
The mass $m$ is due to the interaction with magnetic fields or
magnetic moments perpendicular to the surface 
[Fig.\ \ref{F1}(d)]
and thus breaks time-reversal symmetry.
To study thermal transport, 
we introduce coupling with the fictitious gravitational 
potential
$\phi$
in the Lagrangian \cite{Luttinger1964,Smrcka1977}:
$
{\cal L}=
(1/2)\psi^T\big[i\hbar \partial_t-{\cal H}
-(1/2)\{\phi,{\cal H}\}\big]\psi
$.
The energy-current operator is given by
($a,b=x,y$)
\bea
 j^a_{E}&=&\psi^T \frac{1}{4}\{v^a,{\cal H}\} \psi
\ -\ 
\psi^T\frac{i\hbar}{8}[v^a,v^b]\, \psi
\, \partial_b \phi
\nnn
&&{}
+
\psi^T
\frac{1}{8}
\left[
{\cal H}(v^ax^b+3x^bv^a)+\mathrm{H.c.}
\right]\psi
\, \partial_b \phi.
\eea
The first term is non-vanishing
even in the absence of gravitational potential;
the second and the third terms are proportional to $\nab\phi$, 
which are analogous to the diamagnetic charge current 
in the presence of a magnetic field. 
To evaluate 
$\kappa_{H}=-\langle j_E^x\rangle
/(T\partial_y\phi)
$,
we apply the Kubo linear response formula to the first term, 
while
the second and third terms need to be averaged in thermal
equilibrium \cite{note1}.
When the mass gap is large enough,
the density of states of Majorana fermions vanishes 
at $\varepsilon=0$, and
the thermal Hall conductivity is given by
 $
\kappa_{H}
=
v\, \partial M^z_E/\partial T,
$
where
$M^z_E=(1/4v)\sum_n
f(\varepsilon _n)\varepsilon _n\langle n|(xv^y-yv^x)|n\rangle$,
and $\langle \x|n\rangle=u_n(\x)$ is the exact eigenstate of
the Majorana Hamiltonian (\ref{majorana-hamiltonian}),
${\cal H}u_n(\x)=\varepsilon_nu_n(\x)$.

Quite generally,
one can derive,
in the limit $T\to 0$, 
the relation
\be
 \kappa_{H}
=
\frac{\hbar \pi^2k_B^2T}{6L^2}\sum_{n,m}
\theta(-\epsilon_n)\frac{2{\rm Im}[\langle n|v^x|m\rangle \langle m|v^y|n\rangle]}{(\epsilon_n-\epsilon_m)^2},
\label{WF}
\ee
where $L^2$ is the area of the surface.
Apart from the factor
$\pi^2 k^2_B T/6$,
the right hand side resembles the Kubo formula
for the electrical Hall conductivity,
which, however, is not a well-defined quantity for Majorana fermions;
nevertheless
Eq.\ (\ref{WF}) can be regarded as the generalized Wiedemann-Franz law
to Majorana fermions
\cite{Luttinger1964,Smrcka1977,Vafek2001,Matsumoto2011}.  
Compared to the electron systems, there is an extra factor of 
1/2 due to Majorana nature.
Since
$\sigma_H$ of the massive Dirac fermion is 
$\sigma_{H}^{}= \mathrm{sgn}(m) e^2/(2h)$, 
Eq.\ (\ref{WF}) immediately gives
\bea
\kappa_{H}^{}= \mathrm{sgn}(m) \frac{\pi^2}{6} \frac{k_B^2}{2h}T
\label{kappa-surface}
\eea
for the massive Majorana fermion. 
There is a factor 1/2
compared to the 2d result Eq.\ (\ref{kappa2d}).

\paragraph{Cross-correlated response of 3d-TSC/TSF}

Let us illustrate physical implications of
Eq.\ (\ref{streda-je3}) by studying the responses of
3d-TSCs to the temperature gradient and the rotation.
For simplicity, we consider a sample with cylindrical geometry
with height $\ell$ and radius $r$ as depicted in Fig.\ \ref{F1}(d).
We assume that magnetic impurities are doped near the surface and
their spins are all pointing out or in so that uniform mass gap is
formed on the surface.
Let us first 
introduce the temperature gradient in the $z$-direction,
which generates the energy-current $j_E = \kappa_H \partial_z T$ 
on the surface.
Since $j_E/v^2$ corresponds to the momentum per unit area, 
total momentum due to the surface energy-current is 
$
P_{\varphi}= (2\pi r\ell) j_E/v^2
$
and thus the induced
orbital angular momentum per volume is given by
\begin{align}
\left. L^z \right|^{\ }_{\Omega^z}
=\frac{r P_{\varphi}}{\pi r^2\ell}=
\frac{2}{v^2}\kappa_{H}^{}\partial_zT.
\label{14}
\end{align}
Similarly, upon rotating the cylinder with $\bm{\Omega}=\Omega^z\vz$
(without temperature gradient),
applying Eqs.\ (\ref{streda-je2}) and (\ref{streda-je3}) to the top and
bottom surfaces, we obtain the induced thermal energy density
(the induced entropy change) localized on the top and bottom surfaces, 
\begin{equation}
\left.\Delta Q(z)\right|^{\ }_T=
\frac{2T\Omega^z }{v^2}
\left[
\kappa_H^{\mathrm{t}}\delta(z-\ell/2)
+\kappa_H^{\mathrm{b}}\delta(z+ \ell/2)
\right],
\label{surface Q}
\end{equation}
where $\kappa_H^\mathrm{t}=-\kappa_H^\mathrm{b}$ as the spins on
the top and bottom surfaces are pointing to the opposite directions
(different signs of $m$);
see Fig.~\ref{F1}(d) \cite{Nomura2011}.

In terms of the gravitoelectric field $\vE_g=-T^{-1}\nab T$ 
and
the momentum of the energy current $\vM_E$,
Eq.\ (\ref{14}) can be written as
$
\vM_E= (T\kappa_{H}/v) \vE_g$. 
Further introducing 
the thermal polarization $\vP_E$ by 
$\Delta Q=-\nab\cdot \vP_E$, 
Eq.\ (\ref{surface Q})
can be written similarly as
$\vP_E= (T\kappa_{H}/v) \vB_g$.
These 
highlight
the correspondence between 
TIs and TSCs,
\begin{align}
\mbox{TI:}\, \,
\frac{\partial M^a}{\partial E^b}
=
\frac{\partial P^a}{\partial B^b}
\,\, 
 \Leftrightarrow
\, \, 
\mbox{TSC:}\,\,
\frac{\partial M_{E}^a}{\partial E_{g}^b}
=
\frac{\partial P^a_E}{\partial B_{g}^b}.
\end{align}
Since the orbital angular momentum is given from the internal energy
functional by $L^a=-\delta U_{\theta}/\delta \Omega^a$,
the coupling energy of the temperature gradient and rotation velocity
is written as
\bea
 U_{\theta}=- \int\! d^3\x\, \frac{2}{v^2}\kappa_{H}^{}\nab T\cdot\Vomega
=\int\! d^3\x\, \frac{ k_B^2T^2}{24\hbar v} \theta
\vE_g\cdot\vB_g.\nnn
\label{U_theta}
\eea
This is analogous 
to Eq.\ (\ref{theta_EM}) with
$e^2/\hbar c \leftrightarrow (\pi k_B T)^2/6\hbar v$
and
$\theta=\pm\pi$ playing
the
same role as in Eq.\ (\ref{theta_EM}). 
In 2d-TSC cases, the corresponding term is written as
$
U^{2d}_{\rm TSC}=\int d^2x 
(2/v^2)T\kappa_{H}^{}\phi\Omega^z.
$
This is the
thermodynamical analogue of the Chern-Simons
term \cite{review_TI,Volovik,Volovik2000}.

\paragraph{Angular momentum paradox in 2d-TSC/TSF}
These terms can be related to the problem of total angular momentum of the 
ground state of chiral superfluids.
For simplicity we consider 
the 2d case.
The angular momentum per unit area of the 
ground state of 
the chiral $p$-wave state 
has been proposed to 
behave as
$
L^z(T=0)=-
(\hbar n/2)
(k_BT_c/E_F)^{\gamma}
$
with different exponents 
$\gamma=2$, 1 and 0
\cite{Leggetbook}, 
where $n$ is the number of particles. 
The controversy of $\gamma$ 
has
not yet resolved.
By integrating Eq.\ (\ref{streda-je3}) from 0 to $T$,
and using $\Delta=\hbar vk_F$,
one obtains
$L^z(T)-L^z(0)=
(\pi\hbar k_F^2/6)
(k_BT/\Delta)^2$,
at low temperature, which is consistent with a numerical simulation
\cite{Kita1998}. 
If we extrapolate this relation to the critical temperature $T_c$
at which $L^z(T_c)$ vanishes, we obtain the angular momentum of
the ground state, up to a numerical factor,
$
L^z(T=0)
\sim -\hbar
(\pi k_F^2/2)
(k_BT_c/\Delta)^2
\sim -\hbar n/2,
$
indicative of $\gamma=0$.
(Here we should keep in mind that
we used $\kappa_{H}\propto T$ which is,
however, valid only at low temperature $T\ll T_c$.)
Since the chiral Majorana edge modes of
a 2d-TSC are moving in the opposite direction
to the 
rotation of Cooper pairs {[Fig.\ \ref{F1}(b)],
at finite temperature, 
they contribute to reduce the total angular momentum 
from its ground state value at $T=0$.

\paragraph{Possible experiments}
We conclude by discussing possible experiments to 
probe 
the cross-correlated response of
TSCs.
For the 2d case, let us assume a 2d-TSC is rotating with frequency $\Omega^z$.
As $\Omega^z$ increases 
by $\Delta\Omega^z$, we predict the temperature changes as
$
 \Delta T= \Delta Q/C= 
(2\kappa_{H}T/C v^2)
\Delta\Omega^z,
$
where $C$ is the heat capacity of the 2d-TSC sample.
As the heat capacity of
a fully gapped superconductor
is small
\cite{Maeno2003}, 
this temperature change 
may not be so difficult to measure.

For 3d-TSCs,
the Einstein-de Hass effect 
will reveal
a relationship between magnetism and angular momentum of the system, 
as it has been used for the study of ferromagnetic materials.
We assume a cylindrical 3d-TSC suspended 
by a thin string
and apply thermal gradient [Fig.~\ref{F1}(d)].
This induces surface energy current with angular momentum
$L^z$, according to Eq.\ (\ref{14}).
By the conservation law of total angular momentum,
it must be compensated by a mechanical angular momentum of the material,
which can be directly
measured in principle.
Its inverse effect is the generation of thermal polarization 
(entropy polarization) by rotating the system.
To observe the effect, the frequency $\Omega^z$
should be lower than the critical frequency $\Omega_{c1}$ 
above which vortices are introduced in the bulk of the sample that 
will generate the energy current in the $z$ direction.
The key point is that magnetization has to be induced 
properly on the surface, all pointing out or
in \cite{Nomura2011}.

In summary, we have found nontrivial correlated responses of TSCs to
thermal gradient and mechanical rotation which are analogous to
the topological ME effect.
We have proposed possible
experiments which probe these topological responses.

While we were finalizing the paper, 
a preprint \cite{Qin2011} appeared 
in which the energy magnetization $M_E$ is discussed
while its relation to the angular momentum $L$
and TSCs are not discussed.
This work is
supported by MEXT Grand-in-Aid No. 20740167,
19048008, 19048015, 21244053, Strategic International
Cooperative Program (Joint Research Type) from Japan
Science and Technology Agency, and by the Japan Society
for the Promotion of Science (JSPS) through its ``Funding
Program for World-Leading Innovative R$\&$D on Science
and Technology (FIRST Program).''

\section{thermal Str\v{e}da formula from the CFT partition function}

In this appendix, 
we will derive the thermal  
Str\v{e}da formula (\ref{streda-je3})
of 2d topological liquid
by making use of 
the partition function
of the chiral conformal field theory (CFT) on the edge. 
Our derivation closely follows 
the calculations of 
the thermal conductivity,
the Coulomb blockade current peak, 
and the thermopower
by Cappelli et al
\cite{Cappelli01, Cappelli09, Cappelli09b}.

While the Str\v{e}da formula 
can be derived solely from the Kubo formula in the bulk, 
as illustrated in the text
in terms of the 2d Majorana fermion system, 
the purpose of such complementary derivation is two-fold:
(i) 
In the bulk of a topological fluid, 
the dispersion of gapped quasiparticle excitations 
is not universal and one can adiabatically deform
it without changing the topological properties of the system. 
On the other hand, the dispersion of 
gapless excitations on the edge is,
at low energies, 
quite universal
with emergent Lorentz invariance.

(ii) 
While the derivation in terms of the Kubo formula
mainly focuses,
at least in the form presented in the main text, 
the situation where there is no interaction among
quasiparticles, 
the derivation in term of the edge theory 
applies to
an arbitrary (chiral) topological liquid,
including strongly interacting ones
such as the fractional quantum Hall liquid.

For the Str\v{e}da formula for  
the electrical Hall conductivity in the quantum Hall systems, 
and for the spin Hall conductivity in the quantum spin Hall systems,
an approach 
similar to the calculations we present in this section
can be developed.
Below, we will set
$\hbar= c = k_B =1$.

\subsection{partition function}

Let us consider a topological fluid in 
the annulus geometry
which is boundary by two (inner and outer) edges. 
The low-energy physics is governed by the 
two edge modes. The
Hamiltonian for the edge modes may be written as
\begin{align}
H 
&= 
\frac{v_1}{R_1}
\left(
L_0 - \frac{c}{24}
\right)
+
\frac{v_2}{R_2}
\left(
\bar{L}_0 - \frac{c}{24}
\right),
\end{align}
where $L_0$ and $\bar{L}_0$ represents 
the energy-momentum tensor for the outer and inner edges,
respectively;
$v_{1}$ and $v_2$ are the fermi velocity 
for the outer and inner edges,
respectively;
$R_{1}$ and $R_2$ are the circumference of
the outer and inner edges,
respectively;
By ``adjusting'' the velocities of propagation
of excitations, we may set
$
v_1/R_1
=
v_2/R_2
\equiv
v/R$.

We now consider the grand canonical partition function
for the edges defined by
\begin{align}
Z = \mathrm{Tr}\,  
e^{
- \beta [H 
- \Omega^z L_z
- \mu (N + \bar{N})
- V_o (N-\bar{N})
 ]
}, 
\end{align}
where we have included 
the frequency $\Omega^z$, 
the chemical potential $\mu$,
and the electro potential difference between edges $V_o$;
they couple to 
the total orbital angular momentum,
the total electron number,
and
the difference between the electron numbers in the inner 
and outer edges;
$N$ and $\bar{N}$ represent the total electric charge of
the inner and out edges, respectively.

Below, we will organize these intensive variables 
in terms of two complex numbers
$\tau$ and $\zeta$, 
\begin{align}
2\pi {i} \tau 
=
-\beta \frac{v+{i}\Omega^z/(\pi R)}{R},
\quad
2 \pi {i} \zeta 
=
-\beta (V_o +{i}\mu), 
\end{align}
where we have ``analytically continued''
$\Omega^z \to {i}\Omega^z$,
and
$\mu \to {i}\mu$. 
The grand partition function 
$Z(\tau,\zeta)$
is then given by
\begin{align}
Z(\tau,\zeta)
=
\mathrm{tr}
\left[
e^{ 2\pi {i}\tau (L_0 - c/24)
 -2\pi {i}\bar{\tau} (\bar{L}_0 - c/24)
+\zeta N +\bar{\zeta} \bar{N}}
\right].
\end{align}

In general, for a rational topological phase, 
where the number of types of bulk quasiparticles is finite, 
the annulus partition function can be computed as
\begin{align}
Z(\tau,\zeta)
&=
\sum_{a,\bar{b}}
n^{\ }_{a,\bar{b}}\, 
\chi_{a}(\tau,\zeta)
\overline{\chi^c_{b}(\tau,\zeta)}
\end{align}
where 
$\chi_{a}(\tau,\zeta)$
is a character associated with the quasiparticle of type $a$,
$\overline{\cdots}$ is complex conjugation, 
${\cdots}^c$ represents particle-hole conjugation,
$
n_{a,\bar{b}}
$
is a non-negative integer,
and the summation runs over a finite set of quasiparticle types. 
More specifically, 
the character $\chi_a(\tau,\zeta)$ is given by
\begin{align}
\chi_a(\tau,\zeta) := 
\mathrm{tr}_{a}\, 
\left[
e^{
2\pi{i} \tau (L_0 -c/24)
+
\zeta N
}
\right], 
\end{align}
where $\mathrm{tr}_{a}\left[\cdots \right]$ 
represents the trace within the Hilbert space  
associated to the particle type $a$.

On the disk geometry bounded by only one edge, the partition
function does not consists of 
holomorphic and antiholomorphic parts, 
but reduces to
\begin{align}
Z(\tau,\zeta)
&=
\chi_{a}(\tau,\zeta), 
\label{disk partition function}
\end{align}
where the label $a$ depends on quasiparticles 
present in the bulk. In particular, in the absence of 
quasiparticles, 
we have $a=0$
which represents the identity or ``vacuum'' particle.

\subsection{Str\v{e}da formula
}

In order to compute the thermal Hall conductivity,
$\kappa_{H} = \partial J_{E}/\partial T$, 
we focus on the case with single edge (i.e., disk geometry), 
and compute 
the expectation value of the energy current $J_E$,
which can be written as
\begin{align}
J_E = v \varepsilon
\end{align}
where $\varepsilon$ is the average chiral
energy density on a single edge
and $v$ is the velocity of the edge excitations.
This can be computed from the chiral partition function 
(\ref{disk partition function})
as
\begin{align}
\varepsilon
&=
\frac{{i}v}{2\pi R \beta}
\frac{\partial}{\partial \eta} 
\ln\chi_{a},
\end{align}
where we have introduced $\eta$
\begin{align}
\eta :={i} \Omega^z/(\pi R) 
\end{align}
for notational convenience. 
(Henceforth we set $\zeta=\bar{\zeta}=0$).
The thermal Hall conductivity is then given by
\begin{align}
\kappa_{H}
&= 
\frac{{i}v^2}{2\pi R}
\frac{\partial}{\partial T}
\frac{\partial}{\partial \eta} 
\frac{1}{\beta}
\ln\chi_a.
\label{thermal 1}
\end{align}

On the other hand,
the entropy is give by 
\begin{align}
S &=
\frac{d}{dT} \frac{1}{\beta}\ln Z
=
\frac{d}{dT} \frac{1}{\beta}\ln 
\chi_a.
\end{align}
Taking the derivative with respect to 
$\Omega^z$ on the both sides, 
\begin{align}
\frac{v^2}{2}
\left.
\frac{\partial S}{\partial \Omega^z}
\right|_T
&=
\frac{v^2}{2}
\frac{{i}}{\pi R}
\frac{\partial}{\partial \eta }
\frac{\partial}{\partial T} \frac{1}{\beta}\ln \chi_a.
\label{thermal 2}
\end{align}
We thus obtain
the Str\v{e}da formula for the thermal conductivity:
\begin{align}
\kappa_H
=
\left.
\frac{ \partial J_E}{\partial T} 
\right|_{\Omega^z}
=
\frac{v^2}{2}
\left.
\frac{\partial S}{\partial \Omega^z}
\right|_T.
\end{align}

We now evaluate 
the thermal conductivity
in the large size $R\to \infty$
and
in the low-temperature 
$\beta \to \infty$
limit. It is important to take the $R\to \infty$ limit first,
so we take the limit 
\begin{align}
v\beta/R \ll 1 
\end{align}
first and then take the small $T$ limit.
The limit 
$v\beta/R \to 0$ 
can be evaluated by making use of the modular
transformation of the characters 
\begin{align}
\chi_a(\tau)
=
\sum_b
S_{a}^{b}
\chi_{b}(-1/\tau), 
\end{align}
where $S_a^b$ is the modular $S$-matrix. 
The $v\beta/R\to 0$ limit of the RHS 
can be evaluated as 
\begin{align}
\chi_a(q)
&=
\mathrm{tr}_{a}\, 
\left[
q^{L_0 - c/24}
\right]
\nonumber \\
&
\sim 
-\frac{c}{24} \ln q
+
\ln 
\langle a;0| 
q^{L_0}
|a;0 \rangle 
\nonumber \\
&
=
\left(
-\frac{c}{24} + h_{a}
\right)
\ln q,
\end{align}
where we assume
the modulus of the complex parameter
$q:= e^{-(2\pi {i}/\tau)}$
is small, $|q|\to 0$,
and
$|a;0\rangle$
is the highest weight state,
and
$L_0 |a;0\rangle = h_a |a;0\rangle$. 
Thus,
\begin{align}
\chi_a(\tau)
\sim 
\sum_b
S_{a}^{b}
e^{-(2\pi {i} /\tau)
\left(
-\frac{c}{24} + h_{b}
\right)
}.
\end{align}
In the limit
$
\mathrm{Re}(2\pi {i}/\tau)
=
(4\pi^2 R)/(v\beta)
\to \infty
$,
$b=0$ (the identity particle) dominates in the summation
(unless $S^0_a\neq 0$). Thus,
\begin{align}
\ln \chi_a(\tau)
&
\sim
\ln S_{a}^{0}
+\frac{2\pi {i}}{\tau}\frac{c}{24}.
\end{align}
Extracting the temperature dependence,
\begin{align}
\varepsilon
&\sim 
- 
\frac{{i}}{2\pi R}
\frac{c}{24} 
\frac{\partial}{\partial \gamma}  
\frac{2\pi {i}}{\tau}
= 
\frac{2\pi T^2}{v} 
\frac{c}{24}. 
\end{align}
Finally, $\kappa_{H}$ is given by
\begin{align}
\kappa_{H}
&=
v \frac{\partial}{\partial T}\varepsilon 
= 
v \frac{\partial}{\partial T}
\frac{2\pi T^2}{v} 
\frac{c}{24} 
=
c
\frac{\pi T}{6}. 
\label{calc 1}
\end{align}

Alternatively,
one can obtain
the entropy first:
\begin{align}
S &=
\left(
1- 
\tau
\frac{\partial}{\partial\tau}
\right)
\ln \chi_a
 \nonumber \\
&
\sim
\left(
1- 
\tau
\frac{\partial}{\partial\tau}
\right)
\frac{c}{24}
\frac{2\pi {i}}{\tau}
=
2
\frac{c}{24}
\frac{2\pi {i}}{\tau},
\end{align}
where, again, we take 
the limit $v\beta/R\to 0$. 
The entropy can be expanded in small $\eta$ as
\begin{align}
S
&=
\frac{c \pi^2}{3}
\frac{1}{\beta}
\frac{R}{v}
(1-{i} \eta/v)
+
\mathcal{O}(\eta^2).
\end{align}
We thus conclude
\begin{align}
\kappa_{H}
=
-\frac{v^2}{2}
\frac{1}{\pi R}
\frac{\partial S}{\partial {i}\eta}
=
\frac{c \pi}{6}
T
\end{align}
in agreement with
Eq.\ (\ref{calc 1}).

\newpage

\section{thermal Str\v{e}da formula from the 2d Majonara Hamiltonian}

\subsection{Str\v{e}da formula
}

The energy-current operator is 
divided into
the following two terms:
($a,b=x,y$)
\bea
 j^a_E=j^a_{E\,(0)}+j^a_{E\,(1)},
\eea
where
\bea
 j^a_{E\,(0)}&=&\psi^T \frac{1}{4}\{v^a,{\cal H}\} \psi \\
 j^a_{E\,(1)}&=& - 
\psi^T\frac{i\hbar}{8}[v^a,v^b]\, \psi
\, \partial_b \phi
\nnn
&&{}
+
\psi^T
\frac{1}{8}
\left[
{\cal H}(v^ax^b+3x^bv^a)+\mathrm{H.c.}
\right]\psi
\, \partial_b \phi,
\eea
and $v^a=(i/\hbar)[{\cal H},x^a]$.
$j_{E\,(0)}^a$ is non-vanishing
even in the absence of gravitational potential, while
$j_{E\,(0)}^a$ is proportional to $\nab\phi$.
To evaluate
$\kappa_{H}=-\langle j_E^x\rangle
/(T\partial_y\phi)
$,
we apply the Kubo linear response formula to $j_{E,(0)}^a$:
\begin{align}
-\frac{\langle j^a_{E\,(0)}\rangle}{\partial_b\phi}
&=
-\frac{i}{2L^2}\sum_{nm}\frac{f(E_n)-f(E_m)}{E_n-E_m}
\left(\frac{E_n+E_m}{2}
\right)^2
\nonumber \\
&\quad\quad  
\times 
\frac{\langle n|v^a|m\rangle \langle m|v^b|n\rangle}{E_n-E_m+i\eta}.
\end{align}
Here $\langle \x|n\rangle=u_n(\x)$ is the exact eigenstate of
the Majorana Hamiltonian,
${\cal H}u_n(\x)=\varepsilon_nu_n(\x)$. 
On the other hand, $j_{E (1)}^a$ needs to be averaged in thermal equilibrium:
\bea
 -\langle  j^a_{E\,(1)} \rangle_0
&=&
\frac{1}{2L^2}\sum_n^{}f(E_n)
E_n\langle n|\Big(x^{a}v^{b}-x^{b}v^{a}\Big)|n\rangle \partial_{b}\phi
\nonumber\\
&&+
\frac{1}{2}\frac{1}{L^2}\sum_n^{}f(E_n)
\frac{i\hbar}{4}
\langle n|
[v^{a},v^{b}]|n\rangle \partial_{b}\phi
.
\eea
One obtains the thermal Hall conductivity
\begin{align}
\kappa_{H}
=
\frac{1}{T}
\int_{-\infty}^{\infty}
d\varepsilon 
\left(\frac{-\partial f(\varepsilon )}{\partial \varepsilon }\right)
\frac{\varepsilon ^2}{2}
\int^\varepsilon_{-\infty}
d\zeta\ A^{}(\zeta), 
\label{kappa-1}
\end{align}
where
\cite{Smrcka1977,Vafek2001,Matsumoto2011} 
\begin{align}
A(\zeta)
&=
\frac12 \frac{d B}{d \zeta}
+
\frac{1}{2L^2}{\rm Tr}
\left[
(xv^y-yv^x)\frac{d\delta(\zeta-{\cal H})}{d\zeta}
\right],
\nonumber \\
B^{}
(\zeta)
&=
\frac{i}{L^2}{\rm Tr}\{[v^xG_+(\zeta)v^y-v^yG_-(\zeta)v^x]\delta(\zeta-{\cal H})\},
\label{B-zeta}
\end{align}
and $G_{\pm}(\zeta)=(\zeta-{\cal H}\pm i\eta)^{-1}$.
It is 
convenient to write the thermal conductivity as a sum of 
the following two terms:
\bea
 \kappa_{H}^I 
&=&
\frac{1}{2T}\int\! d\varepsilon 
\frac{\partial f(\varepsilon )}{\partial \varepsilon }\frac{\varepsilon^2}{2}B^{}(\varepsilon ), 
\label{kappa-I}
\\
 \kappa_{H}^{I\!I}\!
&=&
\frac{\partial}{\partial T}\int\! d\varepsilon f(\varepsilon )
\frac{1}{4}{\rm Tr}\Big[
\varepsilon  (xv^y-yv^x)\delta(\varepsilon -{\cal H})\Big] \nnn
&=&
v\frac{\partial M^z_E}{\partial T},
\label{kappa-II}
\eea
where
\bea
M^z_E=\frac{1}{4v}\sum_n
f(\varepsilon _n)\varepsilon _n\langle n|(xv^y-yv^x)|n\rangle.
\eea
At low temperature, 
the Fermi distribution function can be expanded as follows: 
\bea
f(\varepsilon_n-\zeta)=\theta(\zeta-\varepsilon_n)
-\frac{\pi^2}{6}T^2\frac{d}{d\zeta}\delta(\zeta-\varepsilon_n)
\label{f-expand}
\eea
and thus
\bea
\kappa_{H}^I 
\simeq \frac{\pi^2k_B^2T^2}{6} B^{}(\zeta=0).
\eea
When the mass gap 
is large enough, 
the density of states of 
Majorana fermions and Eq.\ (\ref{B-zeta})
vanish at $\varepsilon=0$, 
and hence
$\kappa_{H}^{}=\kappa_{H}^{I\!I}$
while $\kappa_{xx}=0$.

\subsection{Generalized Wiedemann-Franz law
}

Here we derive Eq.\ (\ref{WF}). Noting the identity
\bea
&& \int^{\epsilon}_{-\infty}d\zeta\ A(\zeta) \nnn
&=&\int^{\epsilon}_{-\infty}d\zeta\ \frac{i}{L^2}{\rm Tr}
\Big[
v^x\frac{-1}{(\zeta-{\cal H}+i\eta)^2}v^y\delta(\zeta-{\cal H})-{\rm H.c.}
\Big] \nnn
&=&
\frac{-i}{L^2}\sum_{nm}\theta(\zeta-\epsilon_n)
\left[
\frac{\langle n|v^x|m\rangle\langle m|v^y|n\rangle}{(\epsilon_n-\epsilon_m+i\eta)^2}-{\rm C.c}
\right]
\eea
in Eq.\ (\ref{kappa-1}) and using Eq.\ (\ref{f-expand}),
one obtains Eq.\ (\ref{WF}).


\begin{thebibliography}{10}
\bibitem{TSF} 
We do not distinguish TSFs from TSCs henceforth,
because thermal responses are generic
and common to both of them.
\end{thebibliography}
\end{document}